\documentclass[final,oneeqnum,onefignum]{siamltex} 
\usepackage{epsfig,ifthen}

\title{Stochastic PDEs:
domain formation in dynamic transitions}

\author{
Grant Lythe\footnotemark[1]
}

\begin{document}

\def\volumen{X} \immediate\openin13=volumen.tex
\ifeof13 \else \immediate\read13to\volumen \immediate\closein13 \fi

\immediate\openin14=ultima.tex 
\ifeof14 \else \immediate\read14to\pagina \immediate\closein14
\pagina \stepcounter{page} \fi

\noindent{\small
Proceedings of the VIII Reuni\'on de F\'\i sica Estad\'\i stica,
FISES '97\\ \bf Anales de F\'\i sica, Monograf\'\i as RSEF Vol. 4,
55--63 (1998).}\\[3mm]

\maketitle\label{pini}

\newcommand\mean[1]{{\langle#1\rangle}}
\renewcommand{\thefootnote}{\fnsymbol{footnote}}
\renewcommand{\d}{{\rm d}}
\newcommand{\wt}{{\bf W}_t}
\newcommand{\ws}{{\bf W}_s}
\newcommand{\yt}{{\bf Y}_t}
\newcommand{\real}{{\mathcal R}}
\newcommand{\aaa}{{\mathcal A}}
\newcommand{\borel}{{\cal B}}
\newcommand{\www}{{{\bf w}}}
\newcommand{\eqref}[1]{(\ref{#1})}
\newcommand{\wwt}{{{\bf w}_t}}
\newcommand{\wws}{{{\bf w}_s}}
\newcommand{\yy}{{{\bf Y}}}
\newcommand{\yyy}{{{\bf y}}}
\newcommand{\yyt}{{{\bf y}_t}}
\newcommand{\ee}{{\rm e}}
\renewcommand{\d}{{\rm d}}
\newcommand{\ii}{{\rm i}}
\newcommand{\half}{{\mbox{\(\frac{1}{2}\)}}}
\newcommand{\ha}{\half}
\newcommand{\complex}{{\rm I}\!\!\!{\rm C}}
\newcommand{\pr}[1]{{\cal P}[#1]}

\footnotetext[1]{Centre for Nonlinear Studies,
Los Alamos National Laboratory,
New Mexico NM87545,
USA
}

\renewcommand{\thefootnote}{\arabic{footnote}}

\begin{abstract}
  Spatiotemporal
 evolution in the real Ginzburg-Landau equation is studied
  with space-time noise and a slowly increasing critical parameter.
  Analytical estimates for the characteristic size of the domains
  formed in a slow sweep through the critical point agree with the
  results of finite difference solution of the stochastic PDEs.
\end{abstract}

\pagestyle{myheadings}
\thispagestyle{plain}
\markboth{Grant Lythe}{SPDEs}

\begin{figure} 
\begin{center}
\epsfig{file=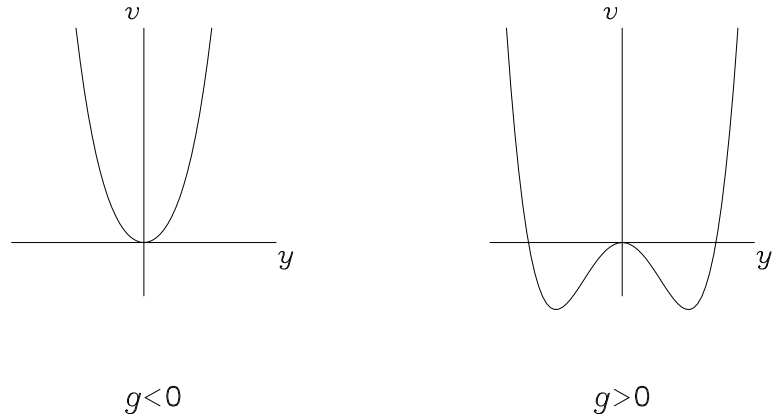,width=5.0in}
\caption[]{
{\em Dynamic transition in terms of a potential}.
The dynamics take the system from 
a single-well potential when the critical parameter $g$
is negative to a double-welled potential for $g$ positive.
When $g$ is explicitly time-dependent, 
the system lingers near the
origin until well after $g=0$. When the increase of $g$
is sufficiently slow, the value of $g$ when the system finally
moves away from the origin is related to the
noise level $\epsilon$ and the rate of increase of $\mu$ by
 $g=\sqrt{2\mu\log\epsilon}$.
}
\end{center}
\label{pot}
\end{figure}

Noise is an indispensable part of the description of symmetry-breaking
transitions, although it is normally included only implicitly in
mathematical models.  This work examines a dynamic transition 
(``quench'') in a
spatially extended system. The model contains a bifurcation parameter
having a critical value at which the state with zero order parameter
becomes unstable to two equivalent states with non-zero order
parameter.  Above this critical value,
the whole system does not choose the same state.  Rather, a
pattern of domains and defects (``kinks'') is formed
\cite{lythedf,candp,landz}.
  Here it is shown that the characteristic
size of the domains formed when the bifurcation parameter is
made explicitly time-dependent depends logarithmically on the magnitude of
additive noise: additive noise is thus a determining influence even
when extremely small. The effect of low-intensity multiplicative noise
is much less dramatic.  Noise is added in such a way that it has no
correlation length of its own (white in space and time).  Numerical
results are reported, obtained using a finite difference algorithm
whose continuum limit is an SPDE.  The simplest model
with the essential features, the real Ginzburg-Landau equation, is
considered here. The dynamics take the system from a
potential with a single minimum to one with two minima
(Figure 1\ref{pot}).
  Similar results have been found for the Swift-Hohenberg
equation, which is more explicitly designed to model Rayleigh-Benard
convection \cite{lythedf}.

\paragraph{The Ginzburg-Landau SPDE}
The order parameter $\yt(x)$ is a function of
 space and time and the bifurcation parameter $g$.  For
$g<0$ fixed, the state $\{\yt(x)=0\ \forall x\}$
 is stable, and only noise
prevents the system from approaching it arbitrarily closely.
For $g>0$ fixed, one sees a pattern of regions in which $\yt(x)$ is
positive and regions in which $\yt(x)$ is negative (domains) separated
by narrow transition layers (kinks). The subject of this paper is not
the slow merging of domains or the nucleation of new kinks - these
happen on timescales much longer than those considered 
here \cite{handl,candp}.  This
paper derives the number of domains formed when the parameter $g$ is
explicitly time-dependent, starting from $g<0$ and ending with $g>0$.

The SPDE describing the dynamics 
is written in the following dimensionless form:
\begin{eqnarray}
\d\yt(x) = (g(t)\yt(x)-\yt(x)^3 + {\cal L} \yt(x))\d t + \epsilon \,\d\wt(x).
\label{parspde}
\end{eqnarray}
Here
$\yt(x):[0,L]^m\times
[-\frac1{\mu},\frac1{\mu}]\times\Omega \to \cal{R}$,
$\Omega$ is a probability space
 and $\wt(x)$ is the Brownian sheet.
The equations were solved as initial value problems, with 
\begin{equation}
g(t)=\mu t
\label{gemt}
\end{equation}
 slowly  increased from $-1$ to $1$.
Periodic boundaries in $x$ are used so that any spatial structure is not a boundary effect.
The constants $\mu$, $\epsilon$ and $\frac1L$
 are all $\ll 1$. The spatial operator
${\cal L}=\Delta$ where
$\Delta=\sum_{i=1}^m\frac{\partial^2}{\partial x_i^2}$, the Laplacian
in ${\cal R}^m$.

An alternative scaling of \eqref{parspde}
is sometimes illuminating: 
if $x$ is rescaled so that $[0,L]\to[0,1]$ then 
\eqref{parspde} becomes
\begin{equation}
\d \yt(x) = (g(t)\yt(x)-\yt(x)^3+D\Delta \yt(x))\,\d t 
+ D^{\frac{m}4}\epsilon \,\d \wt(x),
\label{slspde}
\end{equation}
where $D=L^2$.

\paragraph{The Brownian Sheet}
To add noise to an ordinary differential equation, one adds increments
of the Wiener process at each time step; the idea is of a particle
continuously subject to small impulses \cite{gardiner}.  To add noise
to a partial differential equation (PDE) one adds increments of the
Brownian sheet, that are random in space and time \cite{walsh,pandz}.
One can imagine, for example, the motion of a flexible sheet in a
sandstorm.  Typically the motivation for adding noise to PDEs is to
make allowances for small-scale, rapidly-varying effects with mean
zero that have been neglected in the derivation of an equation.

The Wiener process can be thought of as assigning to each successive
interval of the time axis a Gaussian random variable with variance
proportional to the length of the interval.  The Brownian sheet
assigns to each volume element in $\real_+\times\real^m$ a Gaussian
random variable whose variance is proportional to the volume of the
element \cite{walsh}.  More precisely, it is possible to define a map
$\aaa$ from $\borel(\real_+\times\real^m)$ to a probability space such
that for each $h\in\borel(\real_+\times\real^m)$, $\aaa(h)$ is a
Gaussian random variable with mean zero and $\mean{\aaa(h_1)\aaa(h_2)}
=l(h_1\cap h_2)$ where $l$ is the Lebesgue measure \cite{walsh}.  The
Brownian sheet, so called because its realisations in $m=1$ look like
a ruffled bed-sheet tucked in on two adjacent sides, is defined as
$\wt(x)=\aaa([0,t]\times[0,x])$, where $[0,x]$ is the element
(interval, square, cube, $\ldots$) with opposite corners at the origin
and at $x\in\real^m$. Thus
\begin{equation}
\wt(x):\Omega\times\real_+\times\real^m \quad \to \quad \real.
\label{bsmap}
\end{equation}
The set $\Omega$ is the set of labels for realisations;
averages over realisations are denoted by angled brackets.
  Each
$\wt(x)$ is a real-valued Gaussian random variable with mean zero and
variance
%\begin{equation}
$\mean{\wt^2(x)} = t x^m$.
%\label{varbs}
%\end{equation}
 Stochastic processes will be denoted in bold
with the subscript $t$.  
 For example,  the Wiener process is 
denoted $\wt$ and is constructed  by defining $\wt=\aaa([0,t])$.
 We denote processes
satisfying SPDEs with upper case bold Roman letters
 with subscript $t$ and $x$ in
brackets. 

The action of $\wt(x)$ as an integrator is easily
described in the case where the integrand is a deterministic 
function. If $f_1(x,t)$ and $f_2(x,t)$ are continuous functions on
${\cal{D}}\times [0,t]$, where ${\cal{D}}\in{\cal R}^m$,
 then 
$${\bf I}_t=\int_0^t\int_{\cal{D}}f_1(x,s)\,\d x\,\d \ws(x)
\quad
{\rm and }
\quad
{\bf J}_t=\int_0^t\int_{{\cal{D}}}f_2(x,s)\,\d x\, \d \ws(x)$$
 are Gaussian random variables with 
$<{\bf I}_t>=0$, $<{\bf J}_t>=0$ and 
$$<{\bf I}_t {\bf J}_t>=\int_0^t\int_{\cal{D}}
f_1(x,s)f_2(x,s)\,\d x\, \d s.$$

\paragraph{Numerical Solution}
An approximate numerical solution of an SPDE is 
produced by solving a large set of SDEs.
The finite difference method for a parabolic SPDE consists of
replacing the infinite dimensional system \eqref{parspde} by $N^m$
ordinary SDEs on a grid of equally-spaced points in $[0,L]^m$
separated by $\Delta x$. The SDE at site $i$ is \cite{lythedf}
\begin{equation}
\d \yy_t(i) = (g(t)\yy_t(i)+\yt^3(i))\,\d t 
+ (\Delta x)^{-2}\tilde\Delta \yy_t(i)\,\d t
+(\Delta x)^{-m/2}\epsilon\, \d \wt(i),
\label{fdiff}
\end{equation}
where the discrete Laplacian $\tilde \Delta$ is defined by
\begin{equation}
\tilde\Delta \yy(i) = \sum_{i'}\yy(i')-2m\yy(i)
\label{dislap}
\end{equation}
and the sum is over the $2m$ nearest neighbours of $x$.
 The random variables added at neighbouring grid
points are independent no matter how small $\Delta x$ is. Stochastic
PDEs like \eqref{parspde} use space-time white noise along with a
deterministic operator that includes some spatial coupling (here the
Laplacian) to generate fields with non-delta-function correlations in
space and time. Linear SPDEs can be very efficiently solved
 in Fourier space \cite{gsr,gands}. However,
 the finite difference method is
more easily adapted to nonlinear SPDEs.
One finite-difference numerical realisation of an SPDE
generates an approximate solution for one $\omega\in\Omega$,
at a discrete set of times $\{t_i\}$ and positions $\{x_i\}$.
 The time-stepping used to generate the figures in this
work is the stochastic analogue of the second-order Runge-Kutta method
\cite{gsh,kandp,dynpf}.

\paragraph{Fourier decomposition}
For $k\in{\cal Z}^m$, 
let the Fourier coefficient $\yyy_t(k)$ be defined by
\begin{eqnarray}
\yyy_t{(k)}
 = L^{-\frac{m}2}
\int_{[0,L]^m}\ee^{\ii \kappa x}\yy_t(x)\,\d^m x
\qquad {\rm where}\quad \kappa = \frac{2\pi}{L}k.
\label{fmdef}
\end{eqnarray}
 Each $\yyy_t{(k)}$
 is a complex-valued stochastic process
satisfying the following SDE when $k\ge 1$:
\begin{eqnarray}
\d \yyy_t{(k)} = (g(t)-\kappa^2)\yyy_t{(k)}\,\d t + 
\frac{\epsilon}{\sqrt{2}}\, \d \wt{(k)}
+ \quad {\rm nonlinear}\ {\rm terms}.
\label{fcsde}
\end{eqnarray}
Since $\yt(x)$ is real-valued,
for $k\le -1$ we obtain $\yyy_t{(k)}$ from the relation
$\yyy_t{(-k)}$  = $(\yyy_t{(k)})^*$. 
The imaginary part of $\yyy(0)$ is always $0$;
 the real part satisfies
\begin{eqnarray}
\d \yyy_t{(0,r)} = g(t)\yyy_t{(0,r)}\,\d t + \epsilon\, \d \wt{(0,r)}.
\label{fzsde}
\end{eqnarray}
Each $\wt{(k)}$ is an independent complex-valued Wiener process.
When $|\yyy_t(k)|\ll 1$  for all $k$, the nonlinear terms
that involve products of $\yyy_t(k)$s are unimportant
and the SDEs \eqref{fcsde} can be solved separately.
It then becomes necessary to evaluate the value of $g$
at which these nonlinear terms become important. This
is called an exit value problem in a dynamic bifurcation
\cite{landp1,smm,sha,tsm}.

\paragraph{Dynamic bifurcation}
We consider a  stochastic ordinary
 differential equation slightly more
general than \eqref{fzsde}, including also
 multiplicative noise:
\begin{eqnarray}
\d\yyt = g(t)\yyt\,\d t + \epsilon \,\d
\wwt^{(1)} + \gamma \yyt \,\d\wwt^{(2)},
\label{dbsde}
\end{eqnarray}
where $\wwt^{(1)}$ and $\wwt^{(2)}$ are independent 
real-valued Wiener processes,
$g=\mu t$ and the initial condition is 
$\yyt=y_0$ at $g=\mu t_0=-1$.
$0<\epsilon$, $\gamma\ll\sqrt{\mu}\ll 1$.
 The solution of \eqref{dbsde} is
\begin{eqnarray}
&\yyt=y_0
\exp\left({\frac12\mu(t^2-t_0^2)-\frac12\gamma^2t 
+\gamma\wwt^{(2)}}\right)\nonumber\\ &+ 
\epsilon\, \exp\left({\frac12\mu t^2 - \frac12\gamma^2 t + 
\gamma \wwt^{(2)}}\right)
\int_{t_0}^t\exp\left({-\frac12\mu s^2 + \frac12\gamma^2 s -
 \gamma \wws^{(2)}}\right)\,\d\wws^{(1)}.
\label{dbsoln}
\end{eqnarray}
The second term in \eqref{dbsoln} has zero mean and variance
given by
\begin{equation}
\mean{\yyt^2}-\mean{\yyt}^2
=\epsilon^2 \exp(\mu t^2+\gamma^2 t)
\int_{t_0}^t\exp({-\mu s^2-\gamma^2 t})\,\d s.
\label{meanys}
\end{equation}
For $g>\sqrt{\mu}$, this variance is well approximated by
\begin{equation}
\mean{\yyt^2}-\mean{\yyt}^2\simeq
\epsilon^2\sqrt{\frac{\pi}{\mu}}
{\rm e}^{\frac{1}{\mu}(g+\frac12\gamma^2)^2}.
\label{vary}
\end{equation}
Comparing the mean and standard deviation
of $\yyt$ for $g>\sqrt{\mu}$, we find that,
for $\mu$, $\gamma\ll 1$ and $| y_0|\le {\cal O}(1)$,
the standard deviation is much larger than the mean
if $\mu| \log\epsilon|<\frac12$. Then the dynamic bifurcation
is noise-dominated \cite{landp1},
$\yyt$ remains small until well after $g$ passes through $0$,
and finally attains an ${\cal O}(1)$ magnitude
at a value of $g$ with mean value
$g=g_c+{\cal O}(\mu)$ 
and standard deviation proportional to $\mu$ where
\begin{equation}
g_c=\sqrt{2\mu|\log\epsilon|}.
\label{gcdef}
\end{equation}

Stochastic equations like \eqref{parspde} and \eqref{dbsde} with $g$
constant and $\gamma\ne 0$ have attracted much attention because
multiplicative noise acts in a way that can be interpreted as a shift
of the bifurcation point from $g=0$ \cite{bandk}.
  However, in all cases the shift
is proportional to $\gamma^2$ compared and is dwarfed by the large
dynamic delay of the bifurcation of interest here. In \eqref{vary} we
see that, for the variance of $\yyt$, $g$ is replaced by
$g+\frac12\gamma^2$.  Note, however, that for $\gamma\ne 0$ the
distribution of $\yyt(x)$ is not Gaussian. A slight non-Gaussianity
is the main effect of multiplicative noise in the
SDE and SPDE systems of interest here and will not be further
discussed.

\paragraph{Space-time evolution}
Now return to the spatially extended system.
We first make the hypothesis that, as in the dynamic bifurcation,
$\yt(x)$ remains everywhere small for 
$g<g_c=\sqrt{2\mu|\log\epsilon|}$. Then the emerging pattern 
of domains can be studied from the linearised version
of \eqref{parspde} (i.e. without the cubic term).
The solution of the linearised version of \eqref{parspde} is:
\begin{eqnarray}
\yt(x) = \int_{[0,L]^m}G({ t,t_0,x,v})f(v)\,\d v +
\epsilon\int_{t_0}^t\int_{[0,L]^m}G({ t,s,x,v})\,\d v\,\d\ws(v),
\label{linsoln}
\end{eqnarray}
where $G({t,s,x,v})=$
\begin{eqnarray}
(4\pi(t-s))^{-{m/2}}
\exp\left(-\mu(t^2-s^2)\right)
\sum_{j=-\infty}^{\infty}
\exp\left(-\frac{(x-v-jL)^2}{4(t-s)}\right).
\end{eqnarray}
  The first term in \eqref{linsoln}, dependent on
the initial data $f(x)$, relaxes quickly to very small values.  The
correlation function is therefore obtained from the second,
stochastic, integral in \eqref{linsoln}.
  Performing the integration over space,
assuming that $L>(\frac{8}{\mu})^{\frac12}$, gives
\begin{eqnarray}
c(x)=\mean{\yt(x')\yt(x'+x)}=
\epsilon^2\int_{t_0}^t
\frac{{\rm e}^{\mu(t^2-s^2)}
{\rm e}^{-\frac{x^2}{8(t-s)}}}{(8\pi(t-s))^{\frac{m}2}}\,\d s.
\label{corrint}
\end{eqnarray}

Since $\yt(x)$ satisfies a non-autonomous SPDE,
the correlation function is explicitly a function
of time. In the early part of the evolution, however,
the deviation from that obtained from the corresponding 
static ($g=$constant) equation is small.
We consider this quasi-static period by making the
change of variables $u=t-s$ in \eqref{corrint}.
Then 
\begin{equation}
\exp(\mu(t^2-s^2))=\exp(2\mu t u - \mu u^2)=
\exp(2\mu t u)(1-\mu u^2 +\ldots).
\label{expu}
\end{equation}
and
\begin{eqnarray}
c(x)=\frac{\epsilon^2}{(8\pi)^{m/2}}\left(
\int_0^{\infty}
\frac{{\rm e}^{-2| g| u}}{u^{m/2}}
{\rm  e}^{-\frac{x^2}{8u}}\,\d u
-\mu\int_0^{\infty}
\frac{{\rm e}^{-2| g| u}}{u^{m/2-2}}
{\rm  e}^{-\frac{x^2}{8u}}\,\d u
+\ldots
\right).
\label{corrint2}
\end{eqnarray}
Thus
\begin{equation}
c(x)=\frac{\epsilon^2}2
\frac{\vert g\vert^{\frac{m}4-\frac12}}{(2\pi)^{{m\over 2}}}
x^{1-{m\over 2}}
\left(
K_{\scriptscriptstyle{m\over2}-1}(x\sqrt{{| g|}})
+\frac{\mu}{g^2}\frac{x^2}{16}{\vert g \vert}
K_{\frac{m}2-3}(x\sqrt{{| g|}})
+\ldots\right).
\label{bess}
\end{equation}
where $K_{m}$ is the modified Bessel function of order $m$.
For example, when $m=1$,
\begin{equation}
c(x-x')=
 {\epsilon^2 \over 4\sqrt{| g|}}{\rm e}^{-{{| x-x'|\sqrt{| g|}}}}
\left(1-\frac{3}{16}\frac{\mu}{ g^2}(1+x\sqrt{{\vert g\vert}}
+\frac13x^2{\vert g\vert})+\ldots\right)
.
\label{bess1}
\end{equation}
The first term in \eqref{bess} is the (long-time) correlation
function for the SPDE obtained by fixing $g<0$ \cite{knight}.
Clearly the expansion in $\frac{\mu}{g^2}$ is no longer 
useful for $g>-\sqrt{\mu}$. The correlation
function itself, however,
 remains well-behaved as $g$ passes through $0$;
the divergences associated with critical slowing down
are not present.

In one space dimension, the solution of the SPDE
\eqref{parspde} is a stochastic
process with values in a space of continuous functions 
\cite{walsh,funaki}. That
is, for fixed $\omega\in\Omega$ and $t\in[-\frac1{\mu},\frac{1}{\mu}]$,
one obtains a configuration, $\yt(x)$, that is a continuous function
of $x$. This can be pictured as the shape of a string at time $t$ that
is constantly subject to small random impulses all along its length.
In more than one space dimension, however, the $\yt(x)$ are not
continuous functions but only distributions \cite{walsh,doering} and the
correlation function $c(x)$ diverges at $x=0$.  In the non-autonomous
equations studied here, however, the divergent part does not grow
exponentially for $g>0$, and by $g= \sqrt{2\mu| \log\epsilon |}$
it is only apparent on extremely small scales, beyond the resolution
of any feasible finite difference algorithm.

We now examine the evolution for $g>\sqrt{\mu}$,
where we can approximate \eqref{corrint} using Laplace's
formula:
\begin{eqnarray}c(x) \simeq
\frac{\epsilon^2}{\sqrt{\mu}}
\frac{{\rm e}^{\mu t^2}}{({8 t})^{m/2}}
{\rm  e}^{-\frac{x^2}{8t}}.
\label{rise}
\end{eqnarray} 
Thus typical values of $\yt(x)$
increase exponentially fast and the correlation length 
at time $t$ is  $\sqrt{8t}$. 
Once $c(0)>{\cal O}(\epsilon)$, the noise
no longer greatly influences the evolution; its effect
can be thought of as 
 wiping out the memory of the initial condition at $g<0$
and replacing it with an effective random initial condition.
From \eqref{rise}, we see that, 
  at  $g=\sqrt{2\mu| \log\epsilon |}$,
$c(0)={\cal O}(1)$ and the cubic nonlinearity can no longer be ignored.
(Figure 2.)

\begin{figure} 
\begin{center}
\epsfig{file=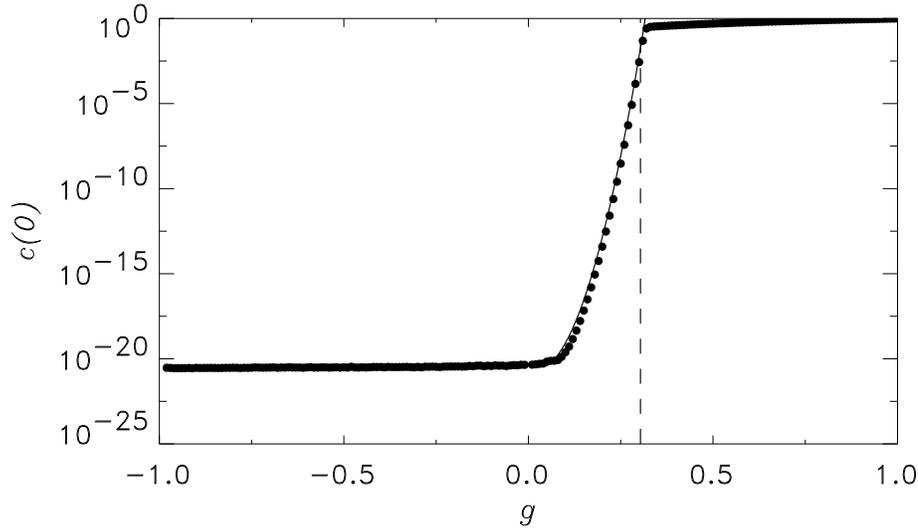,width=5.0in}
\caption[]{
{\em Mean squared value of} \protect$\yt(x)$.
For $g>\sqrt{\mu}$, the variance of
$\yt(x)$ is proportional to $\exp(\mu t^2)$.
This exponential rise eventually takes $\yt(x)$ to
${\cal O}(1)$ values. 
The solid line is \eqref{rise} and the dots are from a
numerical realisation in two space dimensions
with $\mu=0.002$, $\epsilon=10^{-10}$ and $L=12\pi$,
carried out on a $256\times256$ grid.
The vertical dotted line is at $g=g_c=
\sqrt{2\mu| \log\epsilon |}$.
}
\end{center}
\label{risefig}
\end{figure}

\paragraph{Density of kinks}

The second hypothesis that enables the characteristic domain size to
be estimated is that the effect of the cubic nonlinearity, when it
finally makes itself felt, is to freeze in the spatial structure. This
is indeed the case in numerical simulations:
 no perceptible changes occur
between $g=g_c$ and $g=1$ \cite{lythespde,lythedf}.
  Thus the correlation length at $g=g_c$,
\begin{equation}
\lambda=\sqrt{8g_c/\mu}=
2^{7/4}\left(\frac{|\log\epsilon|}{\mu}\right)^{\frac14},
\label{corrl}
\end{equation}
becomes the characteristic length for spatial structure after $g=g_c$.
In one space dimension
 it is possible to put the scenario just described
to quantitative test
by producing numerous realisations 
of the non-autonomous SPDE \eqref{parspde} and recording $r$, 
the number of times that $\yt(x)$ crosses upwards through
$0$ in the domain $[0,L]$ at $g=1$.
The average can then be compared with the mean number of upcrossings
for  Gaussian random field with correlation function \eqref{corrint}
at $g=g_c$.

Consider a homogeneous Gaussian random field $\yt(x)$ 
with correlation function $c(x)$. Then
$\yt(x+\Delta x)-\yt(x)$ is a Gaussian random variable with
mean zero and variance 
$\mean{\left(\yt(x+\Delta x)-\yt(x)\right)^2} = b(\Delta x)$,
where 
\begin{equation}
b(\Delta x) = 2\left(c(0)-c(\Delta x)\right) = 
2c'(0)\Delta x + c''(0)\Delta x^2 +\ldots.
\label{bdef}
\end{equation}
The probability that $\yt$
has an upcrossing of $0$  between
$x$ and $x+\Delta x$ is given by
\begin{eqnarray}
&\pr{({\rm upcrossing}\in (x,x+\Delta x))} \nonumber\\
=&\int_{-\infty}^0\pr{\yt(x)=u}
\pr{\yt(x+\Delta x)-\yt(x)>u}\,\d u \nonumber\\
=&(2\pi)^{-1}({b(\Delta x)c(0)})^{-\frac12}\int_0^{\infty}
\ee^{-u^2/2c(0)}
\int_u^{\infty}
\ee^{-v^2/2b(\Delta x)}\d v\,\d u\nonumber\\
=&\frac12\left(\frac{b(\Delta x)}{\pi c(0)}\right)^{\frac12}
\int_0^{\infty}
\exp(-w^2 \frac{b(\Delta x)}{c(0)})
(1-{\rm erf}(w))\,\d w\nonumber\\
=&
\frac{1}{2\pi}
\arctan\left(\left(\frac{b(\Delta x)}{c(0)}\right)^{\frac12}\right).
\label{probx}
\end{eqnarray}
Now consider a grid of total length $L$ made up of $N$ sites 
separated by $\Delta x$. Let $\Delta x \to 0$ with $L$ fixed, i.e. let
$N\to\infty$.  When $c'(0)\ne 0$, the number of zero crossings 
of $\yt$ defined on this grid is
proportional to $\Delta x^{-\frac12}$ for $\Delta x\to 0$.  When, as
in \eqref{corrint}, $c'(0)=0$, the mean number of zero crossings per
unit length approaches a finite number as $\Delta x\to 0$ given by
\cite{adler,ito}: \begin{equation}
r/L=\frac{1}{2\pi}\sqrt{\frac{-c''(0)}{ c(0)}}.  \label{densz}
\end{equation} In Fig.3 this average is displayed as a function of the
sweep rate for the case \eqref{rise}, {\em evaluated at} $g=g_c$:
\begin{equation} r=\frac{L}{4\pi} \left(\frac{\mu}{2|
\log\epsilon|}\right)^{\frac14}.  \label{meanz} \end{equation}

\begin{figure} 
\begin{center}
\epsfig{file=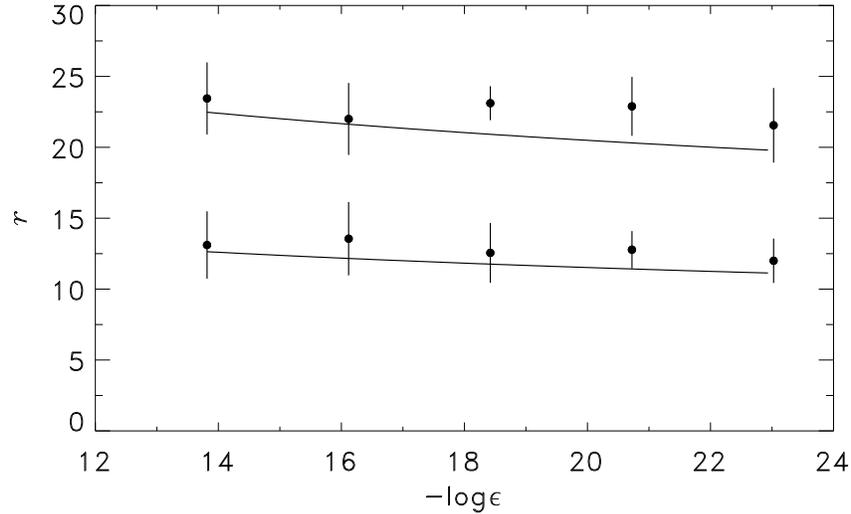,width=5.0in}
\caption[]{
{\em Mean number of zero crossings} at $g=1$.
The solid lines are \protect\eqref{meanz}
with  $L=2048$ and
$\mu=0.01$ (upper line) and $\mu=0.001$ (lower line).
The dots are obtained from finite difference
simulations of \protect\eqref{parspde}.}
\end{center}
\label{xfig}
\end{figure}

\paragraph{Summary}
When the critical parameter in the Ginzburg-Landau equation
 is slowly increased through $0$ 
a characteristic length is produced as follows.
The field remains everywhere small until well after
the critical value for the loss of stability of the 
uniform symmetric state.
Meanwhile, the correlation length
grows proportional to $\sqrt{t}$. 
At $g\simeq g_c=\sqrt{2\mu| \log\epsilon |}$, 
where $\mu$ is the rate of increase of the parameter and
$\epsilon$ is the amplitude of the noise,
the field at last becomes ${\cal O}{(1)}$ 
and the spatial pattern present is frozen in by the nonlinearity.
Thereafter one observes spatial structure with 
characteristic size proportional to
$(\frac{| \log\epsilon|}{\mu})^{\frac14}$.
 The increase of $g$ need not be uniform, but the 
analytical estimates
 presented here rely on the rate of increase
of $g$ as it passes through $0$ being small.

\goodbreak

\label{pend}
\ifthenelse{\isodd{\thepage}}{\newpage\ }{}

\immediate\openout15=nueva.tex
\immediate\write15{\string\setcounter{page}{\thepage}}
\immediate\closeout15
\end{document}